\documentclass[prl,reprint,twocolumn]{revtex4-1}
\usepackage{graphicx}
\usepackage{amsmath, verbatim, amsthm}
\usepackage{mathrsfs}
\usepackage{float}
\usepackage[x11names]{xcolor}
\usepackage{verbatim}
\usepackage{ wasysym }
\usepackage{soul}

\begin{document}

\title{Minimum Action Path theory reveals the details of stochastic biochemical transitions out of oscillatory cellular states}

\author{Roberto de la Cruz} 

\affiliation{Centre de Recerca Matem\`atica, Edifici C, Campus de Bellaterra, 08193 Bellaterra (Barcelona), Spain.}

\affiliation{Departament de Matem\`atiques, Universitat Aut\`onoma de Barcelona, 08193 Bellaterra (Barcelona), Spain.}

\author{Ruben Perez-Carrasco}   
 
\affiliation{Department of Mathematics, University College London, Gower Street, 
London WC1E 6BT, UK} 

\author{Pilar Guerrero}  
 
\affiliation{Department of Mathematics, University College London, Gower Street, 
London WC1E 6BT, UK} 

\author{Tomas Alarcon}\affiliation{ICREA, Pg. Llu\'{\i}s Companys 23, 08010 Barcelona, Spain.}

\affiliation{Centre de Recerca Matem\`atica, Edifici C, Campus de Bellaterra, 08193 Bellaterra (Barcelona), Spain.}

\affiliation{Departament de Matem\`atiques, Universitat Aut\`onoma de Barcelona, 08193 Bellaterra (Barcelona), Spain.}

\affiliation{Barcelona Graduate School of Mathematics (BGSMath), Barcelona, Spain.}

\author{Karen M. Page}  
 
\affiliation{Department of Mathematics, University College London, Gower Street, 
London WC1E 6BT, UK} 

\date{\today}

\begin{abstract} 
Cell state determination is the outcome of intrinsically stochastic biochemical reactions. Transitions between such states are studied as noise-driven escape problems in the chemical species space. Escape can occur via multiple possible multidimensional paths, with probabilities depending non-locally on the noise. Here we characterize the escape from an oscillatory biochemical state by minimizing the Freidlin-Wentzell action, deriving from it the stochastic spiral exit path from the limit cycle. We also use the minimized action to infer the escape time probability density function.  \end{abstract}

\maketitle

\paragraph*{Introduction.}

Cells are intrinsically noisy. Such stochasticity arises not only from the production and degradation of cellular components, but also from their mutual interaction or even the interaction with other cells. Nevertheless, some cellular processes require a precise deterministic output, and noise-suppression mechanisms are necessary within the cell \cite{kepler2001,kaern2005,maheshri2007,losick2008,raj2008}. On the other hand, since fluctuations are an intrinsic component of cellular dynamics, mechanisms are in place that cells exploit to improve its function \cite{cai2008,eldar2010}. For example, randomness can enhance the ability of cells to adapt and increase their fitness in random variable environments \cite{kussell2005,acar2008,guerrero2015a}, or to sustain phenotypic variation \cite{maheshri2007,losick2008,raj2008,macarthur2008,balazsi2011}.

Mean-field descriptions of biochemical processes can be analyzed using dynamical systems theory \cite{distefano2015}, where stable steady states, sustained oscillations or even transients of the ODEs correspond to different possible ceullar states \cite{Jaeger2014,Verd2014}. Relevant examples including sustained oscillations are circadian rhythms \cite{goldbeter1995,leloup2000,smolen2001}, cAMP oscillations in \emph{Dictyostelium} \cite{goldbeter1996}, cell-cycle regulation \cite{gerard2009,gerard2011,gerard2012}, or patterns of bursting in neuronal activity \cite{tsumoto2006,ditlevsen2013,bodova2015,Morris1981,Tateno2004}.
 
When molecular populations are small, the mean-field framework is inaccurate and a stochastic description is required. This involves the formulation of the Master Equation (ME) describing the underlying 
multivariate biochemical birth-death process \cite{anderson2015}. Unfortunately the ME is rarely solvable analytically, requiring the use of Monte Carlo methods (such as the Gillespie algorithm \cite{gillespie1977}). These numerical 
methods are often computationally costly and in usually infeasible \cite{cao2005,cao2005b}. This is especially true in phenomena associated with rare fluctuations, as in the noise-induced escape from a basin of attraction. In these escape problems,  approximations such as the Langevin equation \cite{Kampen2007,gillespie2000} or extreme event theory become necessary \cite{Touchette2009,Perez2016}. In spite of the importance of oscillatory phenomena in biology, most studies have tackled escape problems from point attractors, and a general theory of escape from stable limit cycles is lacking. In order to fill this gap, we consider a simplified oscillatory kinetic model and unveil the ability of the Minimum Action Path (MAP) method from large deviation theory \cite{Freidlin1998,Weinan2004}, which specifies the most probable path between attractors and the mean escape time by minimization of an action functional.

\paragraph*{The model.}

In order to study the limit cycle-fixed point transition we construct a tunable dynamical landscape and then derive the underlying kinetic reactions. This approach allows for a thorough analysis of the escape problem when changing key parameters, such as the angular velocity along the stable limit cycle or the distance between the fixed point and the stable limit cycle.

We construct this prototypical two-dimensional dynamical system, for species X and Y, such that there is a single stable fixed point at $(x_c,y_c)$, and a stable limit cycle of radius $c$ centered at the fixed point (see Fig.\ref{fig.model}). In order to determine the basins of attraction, we include an elliptical repulser, $E(x,y) \equiv 1-\frac{(x-x_c-x_0)^2}{a^2}-\frac{(y-y_c)^2}{b^2}=0$, separating the stable limit cycle from the fixed point. The evolution of the system is given by:  

\begin{figure}
\centering
\includegraphics[width=1.0\columnwidth]{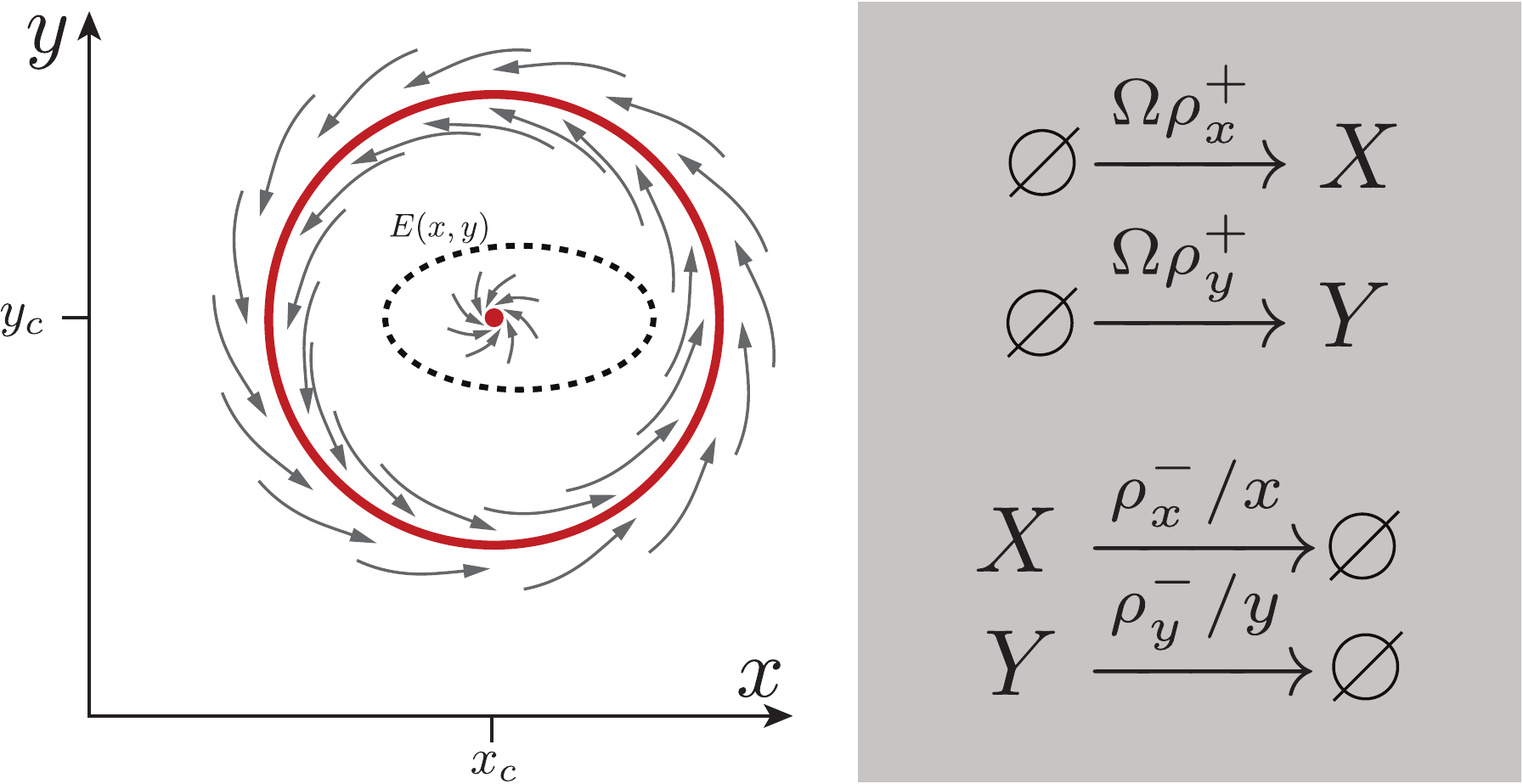}
\caption{\label{fig.model} Left) Dynamical landscape of the model. Stable limit cycle and fixed point are shown in red. Dashed line indicates the repulser $E(x,y)=0$. Right) Underlying set of microscopic biochemical reactions describing the dynamical system.}
\end{figure}

\begin{equation}
\dot{r}  = r (r^2-c^2)E(x,y), \qquad \dot{\theta}  = \omega, \label{eq.model}
\end{equation}
where $r\equiv \sqrt{(x-x_c)^2+(y-y_c)^2}$ is the distance from the fixed point, and $\theta$ is the corresponding angular coordinate, $\tan \theta \equiv (y-y_c)/(x-x_c)$, and $\omega$ is the angular velocity.  Trajectories are either attracted to the fixed point or to the limit cycle, with the exception of 
points lying on an unstable limit cycle, which remain on that orbit. It is important to note that the unstable limit cycle is not the same as the repulser.

\paragraph*{Stochastic description.}

To formulate a stochastic system whose mean-field behavior is given by Eqs. (\ref{eq.model}), we must derive a compatible set of biochemical reactions. Considering four reactions, each species being produced and degraded, the reaction rates are obtained by splitting the rhs of Eq. (\ref{eq.model}) into positive and negative contributions:

\begin{equation}
\dot{x}  = \rho_x^+(x,y) - \rho_x^-(x,y), \quad \dot{y}  = \rho_y^+(x,y) - \rho_y^-(x,y),
\label{eq.modelxy}
\end{equation}

\noindent where:

\begin{equation}
\begin{aligned}
\rho_{x}^+&=&r^2\left(x+x_c(1-E)\right)+c^2(x_c+x(1-E))+\omega y_c,\\
\rho_{x}^-&=&r^2\left(x_c+x(1-E)\right)+c^2(x+x_c(1-E))+\omega y,\\
\rho_{y}^+&=&r^2\left(y+y_c(1-E)\right)+c^2(y_c+y(1-E))+\omega x,\\
\rho_{y}^-&=&r^2\left(y_c+y(1-E)\right)+c^2(y+y_c(1-E))+\omega x_c.\label{eq.rates}
\end{aligned}
\end{equation}

\noindent Identifying $\rho_x^{\pm}$ and $\rho_y^{\pm}$ with the rates of the biochemical reactions, the deterministic system  Eq. (\ref{eq.modelxy})
corresponds to the macroscopic limit of the kinetic reaction set (see Fig. \ref{fig.model}). Here, the system size, $\Omega$, relates the concentrations $x$ and $y$ with the numbers of molecules of each species, $X=x\Omega$ and $Y=y\Omega$. Reactions detailed in Fig. \ref{fig.model} describe a multivariate birth-death process that can be solved numerically using the Gillespie algorithm \cite{gillespie1977}. As $\Omega$ grows, the intrinsic noise is reduced, recovering the mean-field limit (\ref{eq.model}) when $\Omega\rightarrow\infty$. For large (but finite) $\Omega$, the Master Equation can be approximated by the Chemical Langevin Equation (CLE) \cite{gillespie2000},

\begin{equation}
\begin{aligned}
\dot x &=& \rho_x^+ - \rho_x^- + \Omega^{-{1/2}}\sqrt{{\rho_x^+}+{\rho_x^-}} \xi_x(t),\\
\dot y &=& \rho_y^+ - \rho_y^- + \Omega^{-{1/2}}\sqrt{{\rho_y^+}+{\rho_y^-}} \xi_y(t),
\label{eq.CLE}
\end{aligned}
\end{equation}

\noindent where $\xi_x(t)$ and $\xi_y(t)$ are uncorrelated white Gaussian noises, of zero mean, and autocorrelation $\langle\xi_x(t)\xi_x(t')\rangle=\langle\xi_y(t)\xi_y(t')\rangle=\delta(t-t')$. Within this formulation $\Omega$ contributes only to the stochastic terms of the CLE (\ref{eq.CLE}). Therefore tuning the value of $\Omega$ allows us to investigate the role of fluctuations in the transition between the stable limit cycle and the fixed point. 

\paragraph*{Minimum Action Path.}

The intrinsic noise described in the previous section allows for transitions between the limit cycle and the fixed point. Such transitions can occur through many possible transient trajectories, $\varphi(x(t),y(t))$. Nevertheless, not all the transitions are equally probable. In particular, for reaction systems, unlikely transitions decay exponentially with $\Omega$,  $\mathcal{P}\sim \mathrm{e}^{-\Omega \mathscr{S}(\varphi)}$ \cite{Kampen2007,Touchette2009}.  Where the decay rate $\mathscr{S} (\varphi)$ is the so-called action of the transition. This means that for large enough values of $\Omega$, the stochastic transition will concentrate along the path, $\varphi^*$, which minimizes the action:

\begin{equation}
\mathcal{S} \equiv \mathscr{S}(\varphi^*) = \min_{\varphi}\mathscr{S}(\varphi). \label{eq.minaction}
\end{equation}

For the $n$-dimensional stochastic differential equation $\dot{\varphi}=f(\varphi)+g(\varphi)\Omega^{-\frac{1}{2}}\xi(t)$, the action for any path $\varphi_\tau$ of duration $\tau$ is given by the Freidlin-Wentzell functional \cite{Freidlin1998}:

\begin{equation}
\mathscr{S}(\varphi_\tau)=\frac{1}{2}\int_{0}^{\tau} \left\Vert \dot\varphi_\tau(t)-f(\varphi_\tau(t)) \right\Vert^2_{g(\varphi_{\tau}(t))}\mathrm d t, \label{eq.action}
\end{equation}

\noindent where $f(\varphi_\tau)$ is the deterministic field describing the dynamical system, given for our system by the rhs of Eq. (\ref{eq.modelxy}). The multiplicative noise appears in the norm $\Vert\bullet\Vert_{g(\varphi_\tau)}^2$, corresponding with the inner product $\left\langle \bullet,\left(g(\varphi_\tau)g(\varphi_\tau)^\top)\right)^{-1}\bullet  \right\rangle$, where $g(\varphi_{\tau})g^{\top}(\varphi_{\tau})\equiv D$ is the diffusion matrix. Here $D$ takes the form:

\begin{equation}
D(x,y)=\left(\begin{array}{cc} \rho_x^++\rho_x^-&0\\0&\rho_y^++\rho_y^- \end{array}\right).\label{eq.difftens}
\end{equation}

Interestingly, the action and, consequently, the most probable path, are independent of $\Omega$. Additionally, the mean first passage time (MFPT) $T$ from one attractor to the other can be expressed as \cite{Freidlin1998, Touchette2009}:
\begin{equation}
T \simeq C\mathrm{e}^{\Omega \mathcal{S}}. \label{eq.MFPT}
\end{equation} 

In order to find numerically the path minimizing $\mathscr{S}(\varphi_\tau)$, each path of duration $\tau$ was divided into a chain of $N$ segments with initial and final points in the relevant attractors. This reduces finding the optimal path to a minimization problem with $2(N-2)$ degrees of freedom. This was solved using the Broyden-Fletcher-Goldfarb-Shanno algorithm \cite{Fletcher2000,GSL}, using the analytical expression for the gradient of the action in any of the $2(N-2)$ dimensions \cite{Perez2016}.

\paragraph*{Results.}

To assess whether MAP theory can characterize the escape from a stable limit cycle, we have divided the analysis into two sections. First, we compare the MAP with paths obtained numerically from the Master Equation and the CLE. In the second section, we compare MAP predictions of the MFPT with those derived from CLE numerical solutions.

\paragraph*{The MAP predicts average stochastic escape trajectories.}

\begin{figure}
\begin{center}
\includegraphics[width=0.48\textwidth]{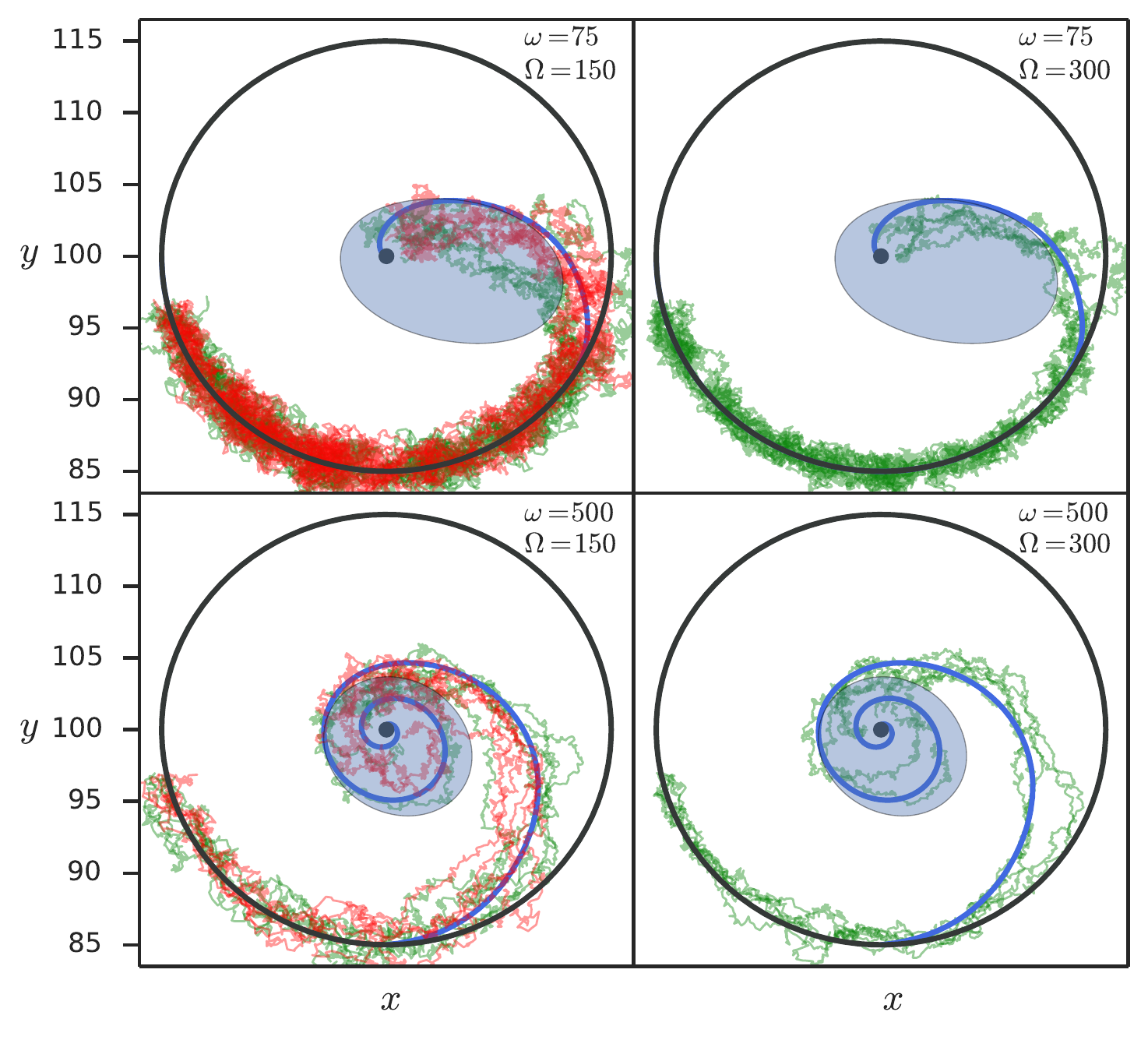}
\caption{\label{fig.scapetrajs}Comparison of escape trajectories from the limit cycle to the stable fixed point. Results show 5 trajectories of the CLE (\emph{green}) compared with the MAP (\emph{blue}) for different values of $\omega$ and $\Omega$. For $\Omega=150$, results are also compared with 5 Gillespie trajectories (\emph{red}). The unstable limit cycle separating the basins of attraction (\emph{shaded area}) is found by temporal inversion of eq. (\ref{eq.modelxy}). For the sake of clarity, only the last turn of each trajectory prior to escape is shown. The rest of the parameters are $x_c$=100, $y_c$=100, $x_0=5$, $a=8$, $b=5$, $c=15$.}
\end{center}
\end{figure}

The MAP defines the most probable transient molecular concentrations during the escape from the stable limit cycle at low noise.  Direct comparison of the MAP with trajectories obtained from numerical integration of the CLE or Gillespie simulations shows good agreement for $\Omega \ge 150$ (Fig. \ref{fig.scapetrajs}). This reveals that, as $\Omega$ increases, the stochastic escape trajectories converge to the MAP. Stochastic simulations for $\Omega<100$ reveal that, when the number of molecular species is low, oscillations have poor quality, and escape trajectories do not concentrate around a single path (data not shown). In addition, our simulations show that the MAP recapitulates changes in escape trajectory with the angular velocity $\omega$ (Fig. \ref{fig.scapetrajs}). 

\begin{figure}
\centering
\includegraphics[width=\columnwidth]{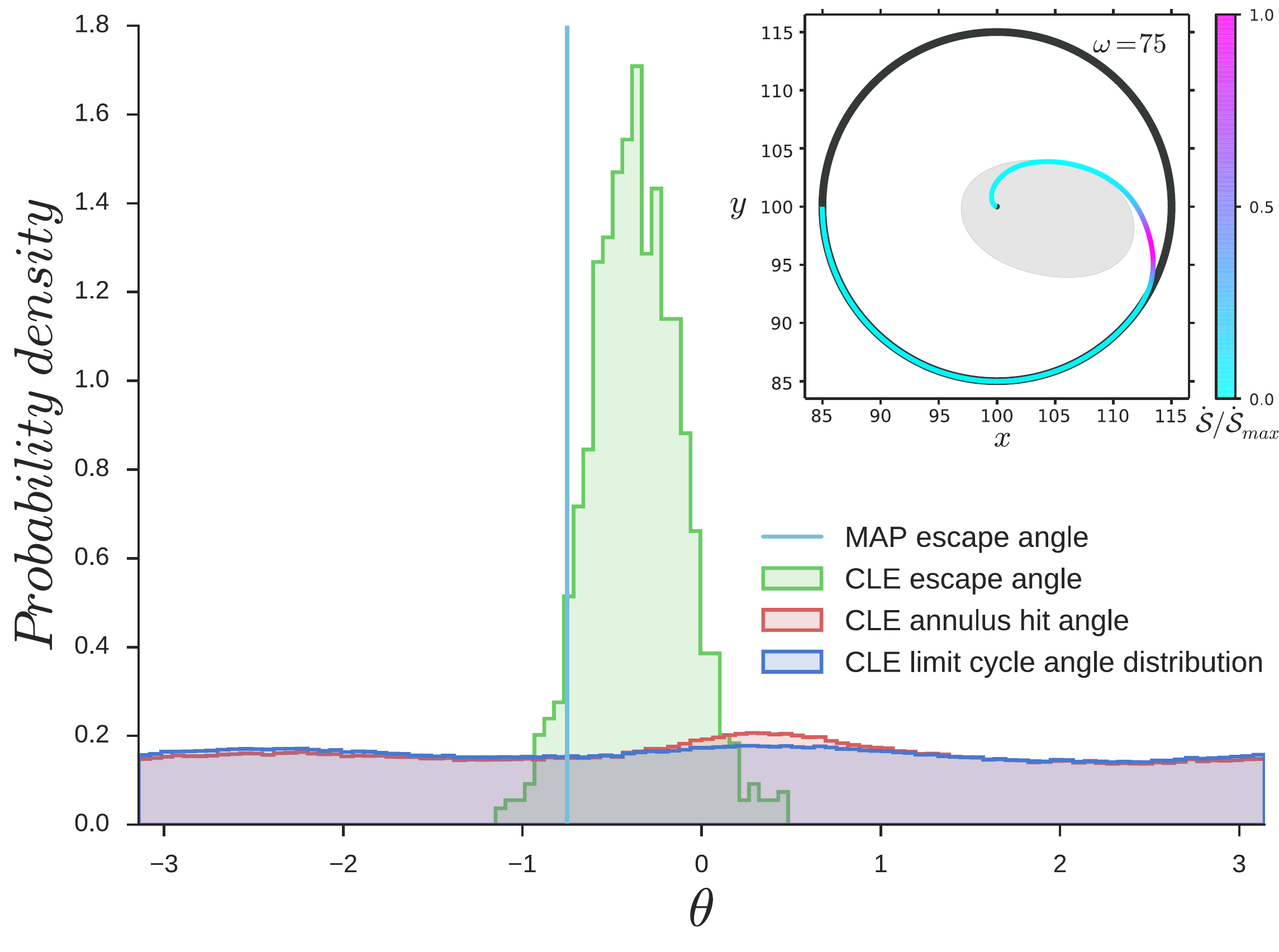}
\caption{\label{fig.escapeangle} Comparison of the escape angle distribution from the limit cycle (\emph{green}) with the escape angle predicted by the 
MAP (\emph{cyan line}) for $\omega=75$ and $\Omega=150$. The results are also compared with the distribution of escape 
angles from an annulus of radius $0.001$ around the limit cycle (\emph{red}), and the angular distribution along the limit cycle, \emph{i.e.} before escape, (\emph{dark blue}). Inset: Action density (Lagrangian) along MAP normalized to the maximum density. 
Other parameters are the same as those of Fig. \ref{fig.scapetrajs}.}
\end{figure}

A more detailed comparison between stochastic simulations and MAP theory reveals that the prediction of the latter becomes less accurate close to the exit point from the cycle. The discrepancy originates in a fraction of trajectories following the limit cycle for a bit longer before starting the transition (Fig. \ref{fig.scapetrajs}).  This results in the prediction of a smaller exit angle than the actual average exit angle (Fig. \ref{fig.escapeangle}). To gain deeper understanding into the origin and magnitude of the discrepancy, we computed the angular probability distribution along the cycle, and the probability distribution of the escape angle from a thin annulus around the limit cycle (Fig. \ref{fig.escapeangle}). The latter has been proposed in \cite{Hitczenko2013} as a quantity that characterizes escape from a stable limit cycle in the low noise limit. Strikingly, our analysis shows that neither measure is as informative as the MAP regarding the escape angle. These results suggest that, even for a simple dynamical system, knowledge of the whole dynamical landscape is required to predict the exit angle from a stable limit cycle, since the purely local analysis around the stable limit cycle does not produce accurate predictions. In this respect, the MAP proves to be useful, since action minimization takes place along the whole escape trajectory. 

Localized inaccuracies in the MAP 
prediction suggest a highly heterogeneous contribution to the action along the MAP. In order to study this, we have evaluated the density of the action 
along the MAP, \emph{i.e.} the Lagrangian of the system. Results show that the density is highest in the middle of the MAP (Fig. 
\ref{fig.escapeangle}), becoming negligible close to the stable and unstable limit cycles, where, in addition, the MAP is tangent to both limit 
cycles. This leads to the discrepancy observed in the exit angles. Note that the portion of the MAP inside the basin of attraction of the stable node does not contribute to the action functional since it corresponds with the deterministic trajectory ($\dot \varphi = f(\varphi(t))$).

In usual escape problems, the path crosses from one basin of attraction to the other at the saddle point of the deterministic system. Here the boundary between basins of attraction is the unstable limit cycle, so the crossing point cannot be identified by a simple local stability analysis. This again shows the predictive power of the MAP approach.

\paragraph*{Minimum action theory predicts MFPT for escape from the cycle.}

\begin{figure}
\includegraphics[width=\columnwidth]{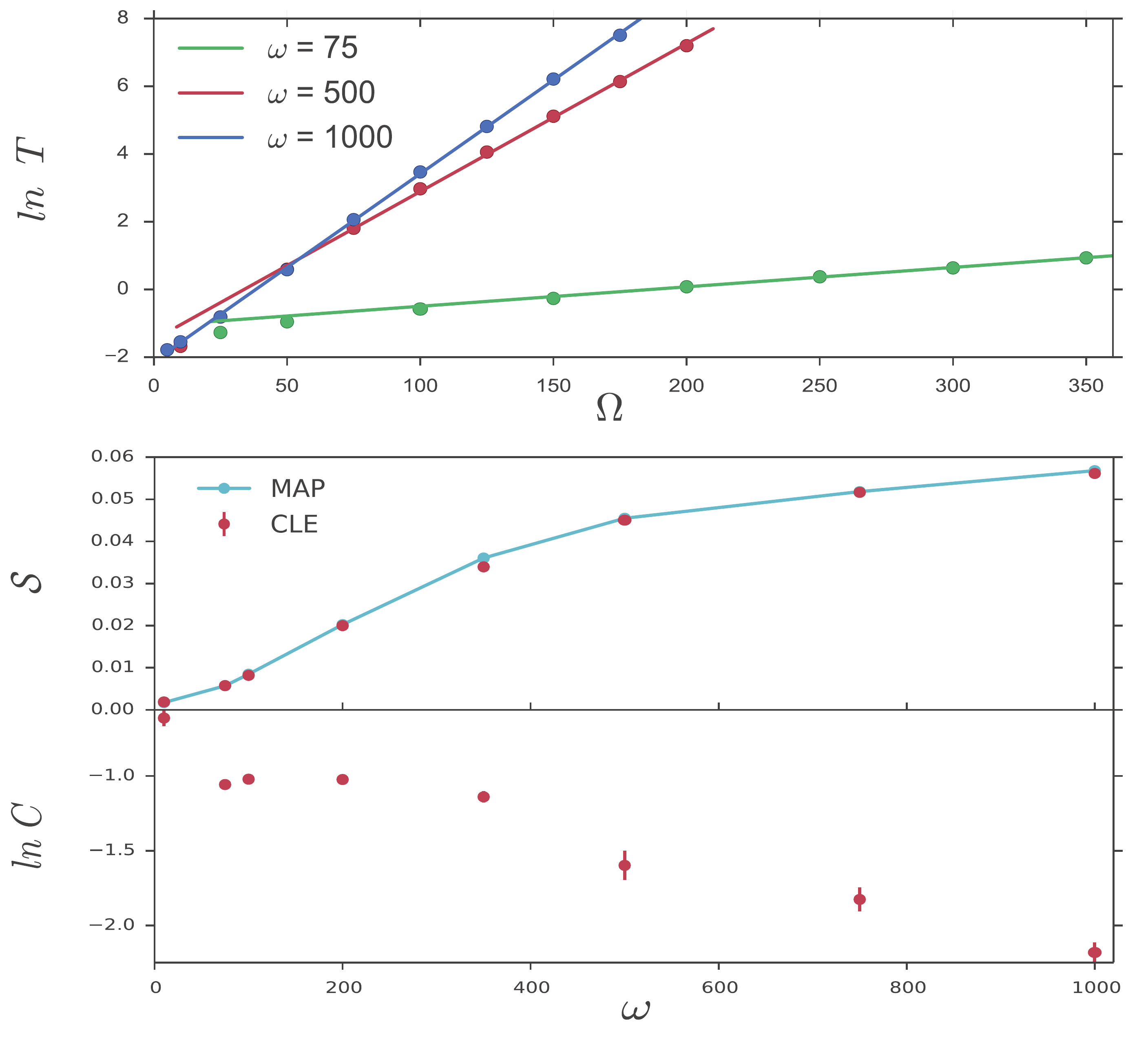}
\caption{\label{fig.MFPT} Comparison of predictions of the MFPT. Top) Comparison of MFPTs calculated from CLE simulations (\emph{circles}) with the exponential dependence of the MFPT on $\Omega$ given by $\mathcal{S}$ (\emph{lines}). Each line is computed by minimizing the action $\mathcal S$ for different $\omega$ and fitting the prefactor $C$. Bottom) Following the same procedure, the value of the action $\mathcal S$ and $C$ are compared for different values of $\omega$.  Parameters values are the same as those of Fig. \ref{fig.scapetrajs}, error bars are standard error of the mean from the CLE.}
\end{figure}

Besides the optimal path, we are also interested in testing the ability of MAP theory to predict the MFPT to exit the basin of attraction. Eq. (\ref{eq.MFPT}) shows that this can be achieved to logarithmic precision, up to a constant, $C$. When the basins of attraction are separated by a saddle point, $C$ can be determined by a Jacobian computed at the saddle  \cite{bouchet2016}. However, in the current case, the separatrix is an unstable limit cycle and $C$ must be computed numerically by solving the CLE at low $\Omega$. It can then be used to predict MFPT for larger $\Omega$, where numerical integration of the CLE is computationally costly. Our results show that the minimum action theory allows us to capture the MFPT dependence on model parameters (Fig. \ref{fig.MFPT}). In our model, we observe an increase in the MFPT with $\omega$. In fact, $C$ also depends non-monotonically on $\omega$ (see Fig.  \ref{fig.MFPT}). Nevertheless, as $\Omega$ grows, the contribution of the prefactor becomes less important ($\ln T \approx \Omega\mathcal S + \ln C$), and the minimum action dominates the escape time estimate.

In addition to the MFPT, we are interested in finding the probability distribution of escape times from the stable limit cycle. Assuming that escape is a rare event focused around a certain exit angle, the escape problem can be described as a Bernoulli process with low success probability $p$ taking place every period of the cycle $\tau=2\pi/\omega$, at times $t_n=2\pi n/\omega$. The probability of exiting at the $n-$th revolution follows the geometric distribution

\begin{equation}
\mathcal{P}(t_n)=p(1-p)^{\frac{\omega t_n}{2\pi}-1}. \label{eq.PDFgeo1}
\end{equation}

Using rare event theory, we can write the success probability as $p=\mathrm{e^{-\mathcal S \Omega}/C}$. The distribution Eq. (\ref{eq.PDFgeo1}) becomes

\begin{equation}
\mathcal{P}(t_n)=\frac{\left(1-\mathrm{e}^{-\mathcal{S}\Omega}/C\right)^\frac{\omega t_n}{2\pi}}{C\mathrm{e}^{\mathcal{S}\Omega}-1}. \label{eq.PDFgeo2}
\end{equation}

Interestingly, since escape is rare, the probability $p$ will be very small and there will usually be many revolutions before the exit from the limit cycle occurs. In the limit $p\rightarrow 0$, the discrete geometric distribution (\ref{eq.PDFgeo2}) can be approximated by its continuum counterpart, the exponential distribution, which does not depend explicitly on the angular velocity,

\begin{equation}
\mathcal{P}(t)= \frac{1}{C}\mathrm{e}^{\mathcal{-S}\Omega+\frac{t}{C}\exp (-\mathcal{S} \Omega)}. \label{eq.PDFexp}
\end{equation}

\begin{figure}	
\includegraphics[width=\columnwidth]{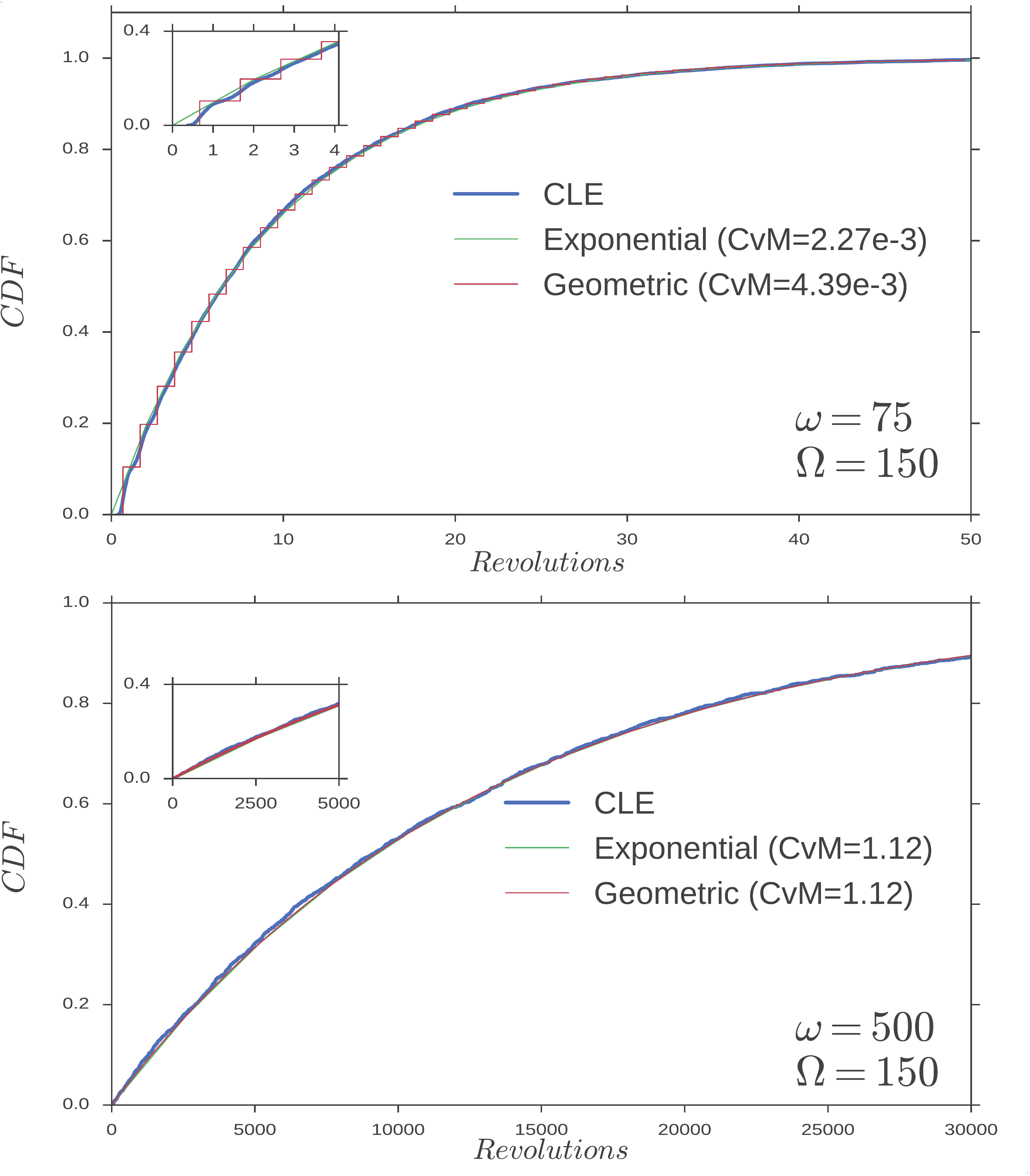}

\caption{\label{fig.CDF} Comparison of the Cumulative Distribution Function (CDF) of the number of turns $t/(2\pi\omega)$ for 10000 realizations with $\omega=75$ and 2000 realizations with $\omega=500$ of CLE. For each distribution the Cramer-von Mises (CvM) criterion is computed to compare the resulting distribution with the geometric and exponential distribution determined by the value of the action. }
\end{figure}

Comparing the distributions Eqs. (\ref{eq.PDFgeo2}) and (\ref{eq.PDFexp}), with the probability distribution of MFPT obtained over several CLE realizations, we obtained a good agreement (see Fig. \ref{fig.CDF}). Surprisingly, even for realizations with a low average number of revolutions prior to escape, the resulting probability distribution is more similar to an exponential distribution than to a geometric one. This is true even for escapes that occur during the first revolution, suggesting that $\theta$ differs significantly from $\omega t$. A more accurate prediction would involve a convolution of geometric processes with the angular noise \cite{Berglund2014, Berglund2016}. However, for the parameters we used, the exponential distribution fits well independently of the average number of revolutions. Fitting the distribution (\ref{eq.PDFexp}) to the MPFTs from CLE realizations therefore provides an alternative method to compute the prefactor $C$.

\paragraph*{Conclusions and perspectives.}

We have shown that, within the rare event theory framework, escape problems from a stable limit cycle can be accurately characterized. The success of the method, in comparison with alternatives, relies on the fact that the Freidlin-Wentzell action is not a local property of the dynamical landscape but of the whole escape trajectory. 
 For sufficiently large system sizes, we have shown that MAP theory accurately predicts escape trajectories and the escape time distribution. The method has also revealed properties of the escape trajectory, such as the tangent exit of the MAP from the stable limit cycle and tangent entry into the basin of attraction of the stable fixed point, as well as the dependence of the entry and exit points on the parameters of the system. 
 
MAP predictions, whilst better than previously used methods, were less successful in determining the exit angle from the stable limit cycle. A deeper analysis of the exit angle could be carried out through a relaxation of the Laplace condition, \emph{i.e.} the reduction to an integral along a single optimal path, by further exploring the distribution of suboptimal paths. Additionally, in the absence of a unique point ( the saddle) separating the basins of attraction, novel research into the calculation of the prefactor \cite{bouchet2016} should be extended.

Eventually, the details of the dynamical landscape will be determined by concrete biological systems. For this reason, future work plans include the application of action minimization to stochastic excitable transitions in type II neurons, where rare event theory will prove its predictive power.  

\begin{acknowledgments}
RdC acknowledges AGAUR-Generalitat de Catalunya for funding under its doctoral scholarship program. RdC and TA are supported by the CERCA Program of the Generalitat de Catalunya, MINECO (grant MTM2015-71509-C2-1-R), and AGAUR  (grant 2014SGR1307). TA acknowledges support from MINECO for funding awarded to the Barcelona Graduate School of Mathematics under the ``Mar\'{\i}a de Maeztu'' program (grant MDM-2014-0445). RP-C, PG and KMP were supported by the Wellcome Trust (grant WT098325MA).
\end{acknowledgments}

%\bibliographystyle{plain} 
%\bibliography{MAPbib}

\end{document}